\documentclass[manuscript]{aastex61}
\usepackage{amsmath,amssymb}
\usepackage{graphicx}
\usepackage{hyperref}

\newcommand{\unit}[1]{\ensuremath{\,\mathrm{#1}}}
\def\kms{$\rm{km~s}^{-1}$}
\newcommand{\bma}[1]{\mathbf{#1}}

\begin{document}

\title{Three-dimensional MHD Simulations of Solar Prominence Oscillations in a Magnetic Flux Rope}
\author{Yu-Hao Zhou}
\affiliation{School of Astronomy and Space Science, Nanjing University, Nanjing 210023, China}
\affiliation{Centre for mathematical Plasma Astrophysics, Department of Mathematics, KU Leuven, Celestijnenlaan 200B, B-3001 Leuven, Belgium}
\affiliation{Key Laboratory of Modern Astronomy \& Astrophysics (Nanjing University), Ministry of Education, Nanjing 210023, China}
\author{C. Xia}
\affiliation{School of Physics and Astronomy, Yunnan University, Kunming 650050, China}
\affiliation{Centre for mathematical Plasma Astrophysics, Department of Mathematics, KU Leuven, Celestijnenlaan 200B, B-3001 Leuven, Belgium}
\author{R. Keppens}
\affiliation{Centre for mathematical Plasma Astrophysics, Department of Mathematics, KU Leuven, Celestijnenlaan 200B, B-3001 Leuven, Belgium}
\author{C. Fang}
\affiliation{School of Astronomy and Space Science, Nanjing University, Nanjing 210023, China}
\affiliation{Key Laboratory of Modern Astronomy \& Astrophysics (Nanjing University), Ministry of Education, Nanjing 210023, China}
\author{P. F. Chen}
\affiliation{School of Astronomy and Space Science, Nanjing University, Nanjing 210023, China}
\affiliation{Key Laboratory of Modern Astronomy \& Astrophysics (Nanjing University), Ministry of Education, Nanjing 210023, China}
\correspondingauthor{C. Xia}
\email{chun.xia@ynu.edu.cn; chenpf@nju.edu.cn}

\begin{abstract}
Solar prominences are subject to all kinds of perturbations during their
lifetime, and frequently demonstrate oscillations. The study of prominence
oscillations provides an alternative way to investigate their internal magnetic
and thermal structures as the oscillation characteristics depend on their
interplay with the solar corona. Prominence oscillations can be classified into
longitudinal and transverse types. We perform three-dimensional ideal
magnetohydrodynamic simulations of prominence oscillations along a magnetic
flux rope, with the aim to compare the oscillation periods with those predicted
by various simplified models and to examine the restoring force. We find
that the longitudinal oscillation has a period of about 49 minutes, which is
in accordance with the pendulum model where the field-aligned component of gravity
serves as the restoring force. In contrast, the horizontal transverse oscillation
has a period of about 10 minutes and the vertical transverse oscillation has a
period of about 14 minutes, and both of them can be nicely fitted with a
two-dimensional slab model. We also find that the magnetic tension force
dominates most of the time in transverse oscillations, except for the first
minute when magnetic pressure overwhelms.
\end{abstract}


\section{Introduction} \label{sec1}
Magnetic field plays an important role in the heating and all kinds of dynamics
of the solar atmosphere. However, the magnetic field in the corona can hardly
be measured directly. Luckily, bodily oscillations of coronal structures such
as coronal loops \citep{asch99, naka99} and solar filaments \citep{trip09},
open a new window to diagnose the corona.
Solar filaments, called prominences when observed above the solar limb (the
two terminologies are used interchangeably in this paper), are cold dense plasma
magnetically suspended in the corona. Their oscillations can provide some clues
to infer the local magnetic structure \citep{arre12}. Compared with coronal
loops, the oscillations of the solar prominences have been studied more
extensively since the discovery of winking filaments \citep{dods49, rams66,
hyde66, klec69}. \citet{rams66} analyzed several oscillating filaments, and one
of them oscillates four times, being triggered by four different flares. The
fact that the periods are the same in the four episodes indicates that the
oscillation period is determined by the intrinsic properties of the filament,
regardless of the origin of the trigger.

Prominence oscillations can be classified in different ways. Early on,
prominence oscillations were classified into short- and long-period
oscillations, with the periods being $\le$10 min \citep[e.g.][]{tsub86} for the
former and $\ge$40 min for the latter \citep[e.g.,][]{bash83, bash84}. It is
noted that oscillations with short and long periods can co-exist in one
prominence \citep{bocc11}. Recent
studies usually classified them depending on the velocity amplitudes into
small-amplitude oscillations \citep[$\sim$2--3 \kms, see][for a review]{oliv02,
oliv09, arre12} and large-amplitude oscillations \citep[$\ge$20 \kms, see][for
a review]{trip09}. Another widely used classification is based on the
oscillation direction relative to the magnetic field. In this case, prominence
oscillations can be divided into longitudinal oscillations, whose direction of
motion is parallel to the local magnetic field inferred by the filament
threads, and transverse oscillations, where the displacements are perpendicular
to the filament threads. On average, longitudinal oscillations show a longer
period than transverse oscillations. This kind of classification might be more
physical since it seems that longitudinal and transverse oscillations have
their own individual restoring forces, which are the crucial factor for oscillations.
However, it should be emphasized here that it is not straightforward to
distinguish the longitudinal and transverse modes: whether an oscillation is
longitudinal or transverse is determined by the oscillation direction relative
to the local magnetic field (or filament threads), not relative to the
filament spine. Since the filament threads are skewed from the filament spine
with an angle of several to 30 degrees \citep{atha83, hana17}, longitudinal
oscillations would also manifest transverse displacement relative to the
filament spine in H$\alpha$ images \citep{pant15, chen17}. Note that the transverse
oscillations can be further divided into horizontal ones \citep{isob06,
asai12, gosa12, shen14} and vertical ones \citep{eto02, okam04, gilb08}.

Transverse oscillations of prominences with a typical period of 10--20 minutes
have been theoretically investigated
since 1960s. In early studies, \citet{hyde66} and \citet{klec69} approximated
an oscillating prominence as a single mass harmonic oscillator, with the
magnetic tension force being the restoring force. Later the prominence was
modeled as a dense cold slab embedded in the hot tenuous corona along a
magnetic flux tube \citep[see][for an example]{joar92}, where the slab has a
finite length but infinite width and height. In such a slab model,
the
global oscillation of the prominence is described as the string model, with the
oscillation period determined by
\begin{equation}
  P=2\pi(WL)^{1/2}/v,
  \label{eq11}
\end{equation}
where $L$ is the half length of the flux tube, $W$ is the half length of the
prominence, while $v$ represents a typical fast, slow or Alfv\'en speed in the
prominence. Following these pioneering explorations, more complicated models
for transverse oscillations were proposed. In these models, more observational
facts are considered, such as gravity \citep{oliv93}, the angle between the
prominence and magnetic lines \citep{joar93}, the prominence-corona transition
region \citep[PCTR, ][]{oliv96}, finite transverse extension of the prominence
\citep{diaz01, diaz02}, non-adiabatic effects \citep{terr01}, mass flows
\citep{sole08}, ion-neutral collisions \citep{sole10}, and so on \citep[see]
[for a review]{arre12}. It is noticed that the increased complexity usually
fails to allow for a simple analytical solution. Moreover, the magnetic flux
tube which supports the prominence has a 3-dimensional (3D) nature with strong
curvature. As a result, the vertical and horizontal transverse oscillations
might display significant differences. However, in various simplified models,
horizontal and vertical transverse oscillations are identical. All these
features can be better captured with 3D magnetohydrodynamic (MHD) simulations,
and the results in the above-mentioned analytical models can be examined or
verified with 3D MHD simulations.

Longitudinal oscillations were discovered only 15 years ago \citep{jing03}.
Since then, many cases have been reported \citep{jing06, vrsn07, li12, zhan12,
luna14, shen14, zhan17}. This type of oscillations has a period of the order
of 1 hour, several times longer than that of typical transverse oscillations.
\citet{jing03} proposed several mechanisms of the restoring force to explain
the longitudinal oscillations of filaments, such as field-aligned gravity and
pressure enhancement due to an Alfv\'en wave package bouncing back and forth
along the anchored magnetic loop. They also considered a possibility that the
observed longitudinal oscillations might be an apparent motion due to
successive transverse oscillations of neighboring threads of the filament.
\citet{vrsn07} suggested the magnetic pressure gradient to be the restoring
force of the longitudinal oscillations, where the increase of magnetic pressure
was thought to result from the injection of poloidal magnetic flux into the
filament via magnetic reconnection. In their model, the oscillation period $P$
was derived to be expressed as $P\approx 4.4 L/v_{A\phi}$, where $L$ is the
half length of the magnetic flux rope and $v_{A\phi}$ is the Alfv\'en speed
based on the equilibrium poloidal field of the filament.

Later, more efforts were spent on the field-aligned component of gravity as the
restoring force. For example, \citet{luna12a} proposed the ``pendulum model" to
explain the filament longitudinal oscillations, where the field-aligned
component of gravity serves as the restoring force for the filament threads to
oscillate along the magnetic dips. With an analogy to the pendulum, the
oscillation period is determined by the curvature radius ($R$) of the magnetic
dip, i.e., $P=2\pi\sqrt{R/g}$, where $g$ is the solar gravitational
acceleration near the solar surface. More convincingly, \citet{zhan12} compared
1-dimensional (1D) radiative hydrodynamic numerical simulations with
observations of prominence longitudinal oscillations. In their simulation
setup, the geometry of the magnetic dip, which determines the curvature radius
$R$, was taken from observations. It turned out that the oscillation period in
the simulation is consistent with the observations. Since they found in their
simulations that the gravity component overwhelms the gas pressure gradient,
their results strongly favor the field-aligned component of gravity as the
restoring force for filament longitudinal oscillations. They further performed
a parameter survey on how the oscillation period and decay time depend on the
geometry of the magnetic configuration.

\citet{terr13} extended the simulations into two-dimensions (2D) by numerically
solving the linearized MHD equations. However, they found that the oscillation
period is 2--3 times larger than that predicted by the pendulum model. Regarding
this discrepancy, \citet{luna16} pointed out that the inconsistency is
due to the fact that the filament in \citet{terr13} is supported by magnetic flux tubes
with too shallow dips. If the magnetic dips are too shallow, the field-aligned
component of gravity no longer overwhelms the gas pressure gradient. When the
filament is located in deeper magnetic dips, further simulations indicate that
the oscillation period becomes consistent with the pendulum model again. So
far, the filament longitudinal oscillations were simulated in 1D and 2D only,
where the magnetic configuration is remarkedly simplified. To have a more
realistic magnetic configuration, we need to resort to 3D MHD simulations.

In this paper, we aim to perform 3D MHD simulations of both longitudinal and
transverse oscillations of solar filaments. Our paper is organized as follows.
The setup of our simulation and the numerical method are described in Section
\ref{sec2}. The numerical results of the simulations are presented in
Section \ref{sec3}, which is followed by discussions in Section \ref{sec4}.

\section{Numerical Method} \label{sec2}

Prominence oscillations are observed in both active region prominences and
quiescent prominences, and we know that active region prominences usually
have a stronger magnetic field than quiescent prominences. Since a strong
magnetic field implies the need for more computational resources, a model
representing quiescent prominences is selected for our simulation. Statistical
analysis reveals that $\sim$96\% of the quiescent prominences are supported by
a magnetic flux rope \citep{ouya17}. Therefore, a flux rope is adopted as the
magnetic structure for our simulations.

Our basic setup is similar to \cite{xia16}. We start from a static coronal
volume in a Cartesian box. The box extends in $-180$ Mm $< x <$ 180 Mm, $-120$
Mm $< y <$ 120 Mm, and 0 $< z <$ 240 Mm. The number density starts from
$10^{9} \unit{cm}^{-3}$ at the bottom boundary and decreases exponentially to
satisfy the hydrostatic equilibrium in a 1 MK isothermal corona. In order to
obtain a force-free magnetic field, we prescribe the following distribution of
the $z$-component of the magnetic field in a plane below our bottom boundary at
$z=-4$ Mm, where the $z$-axis is upward:
\begin{equation}
    {B_{z}}(x,y)=
	\begin{cases}
        0 & y < - {\delta _y}; \\
		{B_{z0}}\sin (\pi y/{\delta _y})\exp (x_m^2/\delta _x^2) & - {\delta _y} \le y \le {\delta _y};\\
		0 & y > {\delta _y};\\
	\end{cases}
    \label{eq21}
\end{equation}
Here $x_m = \rm{minmod}$${(x+x_0, x-x_0)}$ is the median of $x+x_0$, $x-x_0$, and
0. The parameters in Equation (\ref{eq21}) are chosen as follows: $B_{z0}$ = 25
G, $\delta_x$ = 30 Mm, $\delta_y$ = 80 Mm and $x_0$ = 50 Mm. The bipolar
magnetogram described by Equation (\ref{eq21}) is placed below our bottom
boundary in order to avoid any sharp variation of magnetic field resulting from
the extrapolation. The force-free parameter $\alpha$ in our extrapolation is
chosen to be a constant, $-0.08$. The resulting plasma $\beta$ ranges from
0.015 to 0.5 for $z < 100$ Mm, and goes up to about 1.4 near the top boundary.

In order to form a magnetic flux rope from the above-mentioned force-free
field, we first perform simulations by solving the following isothermal MHD
equations:
\begin{equation}
    \frac{{\partial \rho }}{{\partial t}} + \nabla  \cdot (\rho \bma{v}) = 0,
    \label{eq22}
\end{equation}
\begin{equation}
    \frac{{\partial (\rho \bma{v})}}{{\partial t}} + \nabla  \cdot (\rho \bma{vv} + {p_{tot}}\bma{I} - \frac{{\bma{BB}}}{{{\mu _0}}}) = \rho \bma{g},
    \label{eq23}
\end{equation}
\begin{equation}
    \frac{{\partial {\bma{B}}}}{{\partial t}} + \nabla  \cdot (\bma{vB - Bv}) = 0,
    \label{eq24}
\end{equation}
where ${p_{tot}}=p+{B^2}/2{\mu _0}$ is the total pressure, $\bma{g}=-{g_\odot}
{r_\odot}^2/{({r_\odot}+z)^2}{\mathbf{\hat e}_z}$ is the gravitational
acceleration, and ${g_ \odot} = 274 \unit{m}\unit{s^{-2}}$ is the gravitational
acceleration at the solar surface, and ${r_\odot}=691 \unit{Mm}$ is the solar
radius. All the other symbols in the equations have their usual meanings. The
evolution is driven by a surface flow that is described as follows:
\begin{equation}
	\begin{cases}
        {v_x}(x,y) = f(t)C(\partial|{B_{mz}}|/\partial y)\exp
             (-{y^2}/{\delta_y}^2)[{\mathop{\rm sgn}}(y+{\delta_y}/2)-
             {\mathop{\rm sgn}} (y - {\delta_y}/2)]; \\
        {v_y}(x,y)=-{v_x}(x,y);\\
        {v_z}(x,y)=0,\\
    \end{cases}
    \label{eq25}
\end{equation}
where $t$ is the time and $f(t)$ is a linear ramp function allowing to
progressively change the driving velocity according to
\begin{equation}
    f(t)=
	\begin{cases}
        t/t_{ramp} & 0 \le t < t_{ramp}; \\
        1 & t_{ramp} \le t \le t_{max}-t_{ramp}; \\
        (t_{max}-t)/t_{ramp} & t_{max}-t_{ramp} < t \le t_{max}. \\
    \end{cases}
    \label{eq26}
\end{equation}
In our simulation, $t_{ramp}$ and $t_{max}$ are 14.3 min and 100.2 min,
respectively. The parameter $C$ is used to control the maximum value of our
driving velocity to be 12.8 \kms, which is larger than observational values,
but is still much smaller than the Alfv\'en speed in the corona.

The normal component of the magnetic field at the boundaries is derived from
the inner points in order to keep the field divergence-free (in a centered difference
scheme). For other variables in the four lateral boundaries, a zero-gradient
extrapolation is applied. At the bottom boundary, density is fixed to keep the
gravitational stratification. At the top boundary, we extrapolate the velocity
and adopt a gravitationally stratified density profile.

\begin{figure*}
	\centering
	\includegraphics[width=\linewidth]{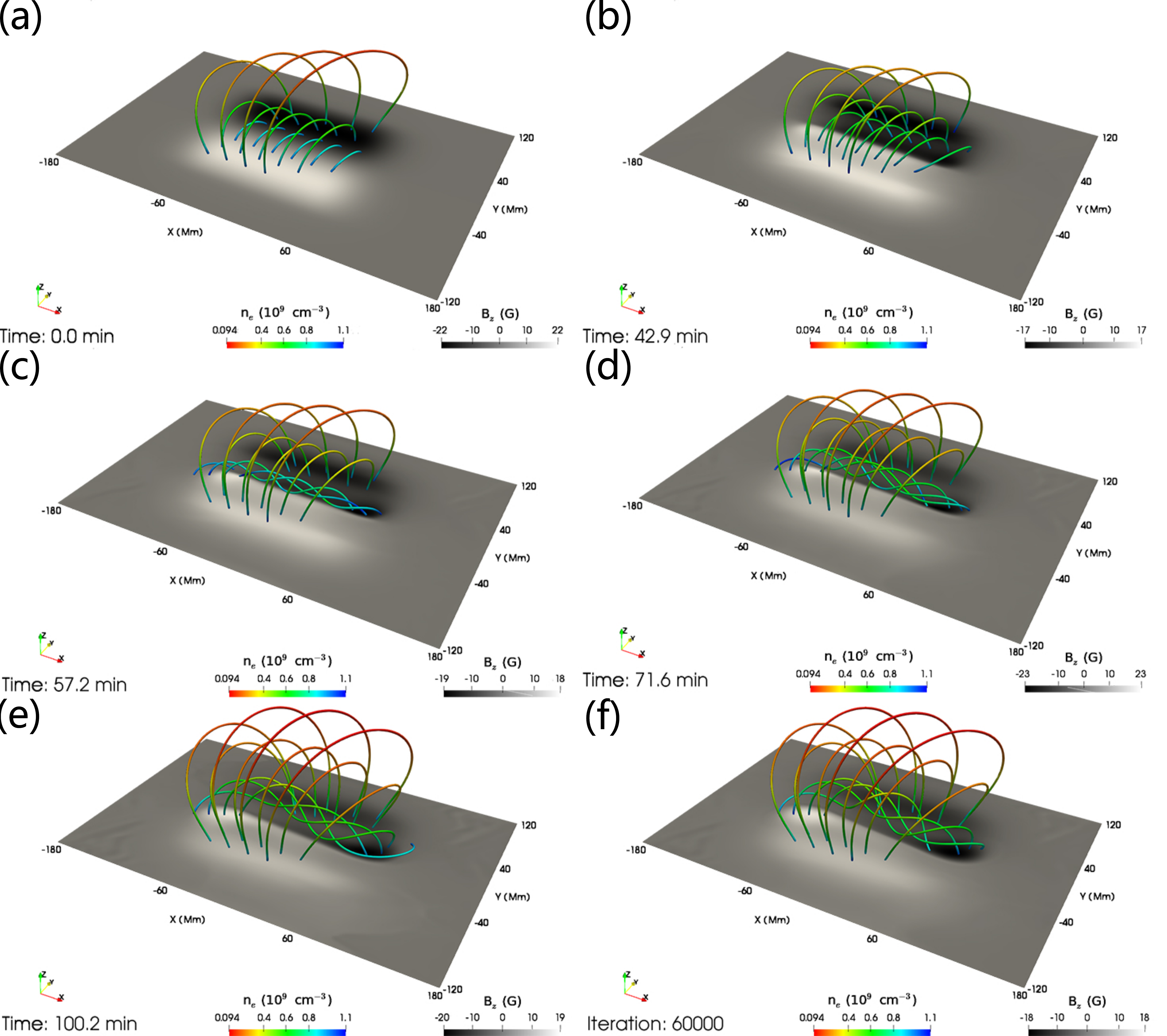}
	\caption{Panels (a--e) show five moments of the magnetic field evolving from a sheared arcade to an elongated flux rope. The field lines are colored by number density. The grayscale in the bottom plane indicates the evolving $z$-component of the magnetic field. Panel (f) shows the magnetic field we get after the magneto-frictional relaxation.}
    \label{fig1}
\end{figure*}

Equations (\ref{eq22}--\ref{eq24}) are numerically solved using the adaptive
mesh refinement (AMR) versatile advection code \citep[MPI-AMRVAC 2.0,][]{xia18,kepp12,
port14}. A four-level AMR grid is used, whose base grid level is $144 \times 96
\times 96$ and it will reach an effective resolving power of $312 \unit{km}
\times 312 \unit{km} \times 312 \unit{km}$ at the finest cells. As shown in
panels (a--e) of Figure \ref{fig1}, after imposing the driving flow, the
magnetic field becomes more and more sheared. After about 50 minutes, a small
flux rope is formed, and then it grows while rising slightly. Then, after
another 50 minutes when the driving flow is completely stopped, we get a large
elongated flux rope. This flux rope has a length of about 200 Mm in the
$x$-direction with a diameter of its cross section of about 40 Mm. The centre
of the flux rope is located at a height about 35 Mm from the bottom boundary
and the maximum magnetic field strength is about 16 G.

However, we found that the flux rope formed this way is not force-free enough.
Therefore, at the end of this stage, a magneto-frictional method is imposed for
60,000 iteration steps \citep[see][for details of this method used in
MPI-AMRVAC]{guo16}. Figure \ref{fig1}(f) shows the magnetic field lines we
eventually got. While apparently the configuration does not change too much
compared to Figure \ref{fig1}(e), actually the maximum current density is
reduced by half, which implies that the magnetic field becomes smoother. For
prominence longitudinal oscillations, the magnetic field we then obtain is fairly weak
so that also transverse oscillations would be easily excited. In order to avoid
such mode coupling, we multiply the magnetic field by a factor of 1.5 for the
simulation of longitudinal oscillations. Since each component of the magnetic
field is amplified by the same factor, the resulting magnetic field is
force-free as well.

So far, we have obtained a hydrostatic isothermal atmosphere and an almost
force-free magnetic field with a 3D flux rope embedded in an envelope field. As
the final step to get our initial setup for prominence oscillations, we follow
\citet{xia16} and replace the isothermal atmosphere with an idealized
chromosphere and corona, whose temperature distribution is expressed as
follows:
\begin{equation}
    T(z)=
	\begin{cases}
        {T_{ch}} + ({T_{co}} - {T_{ch}})(1 + \tanh (z - {h_{tr}} - {c_1})/{w_{tr}})/2 & z \le h_{tr}, \\
        {(7{F_c}(z - {h_{tr}})/(2\kappa ) + {T_{tr}}^{7/2})^{2/7}} & z > h_{tr}, \\
    \end{cases}
    \label{eq28}
\end{equation}
where $h_{tr}=4$ Mm is the height of our `transition region'. The transition
region is slightly higher than in reality \citep[see also][]{hill13, hans17}.
We take $T_{ch}=1.5\times 10^4$ K, $T_{tr}=1.6\times 10^5$ K and $T_{co}=1.5
\times 10^6$ K, which are typical values for the temperatures of the
chromosphere, the transition region, and the corona, respectively. $F_c=2\times
10^5\unit{erg} \unit{cm}^{-2} \unit{s}^{-1}$ is the constant vertical thermal conduction
flux and $\kappa=10^{-6}{T^{5/2}} \unit{erg} \unit{cm}^{-1}
\unit{s}^{-1} \unit{K}^{-1}$ is the Spitzer-type heat conductivity. Then, we
use a hyperbolic tangent function with parameters $c_1=0.333$ Mm and $w_{tr}=
0.3$ Mm to extend the temperature profile from the corona into the
chromosphere. The resulting temperature ranges from $1.5\times 10^4$ K at the
bottom boundary to about $2.3\times 10^6$ K near the top boundary. By assuming
a hydrostatic atmosphere, we then derive the density distribution $\rho_{old}$,
starting from a given number density of $8.33\times 10^{12} \unit{cm}^{-3}$ at
the bottom.

\begin{figure*}
	\centering
	\includegraphics[width=\linewidth]{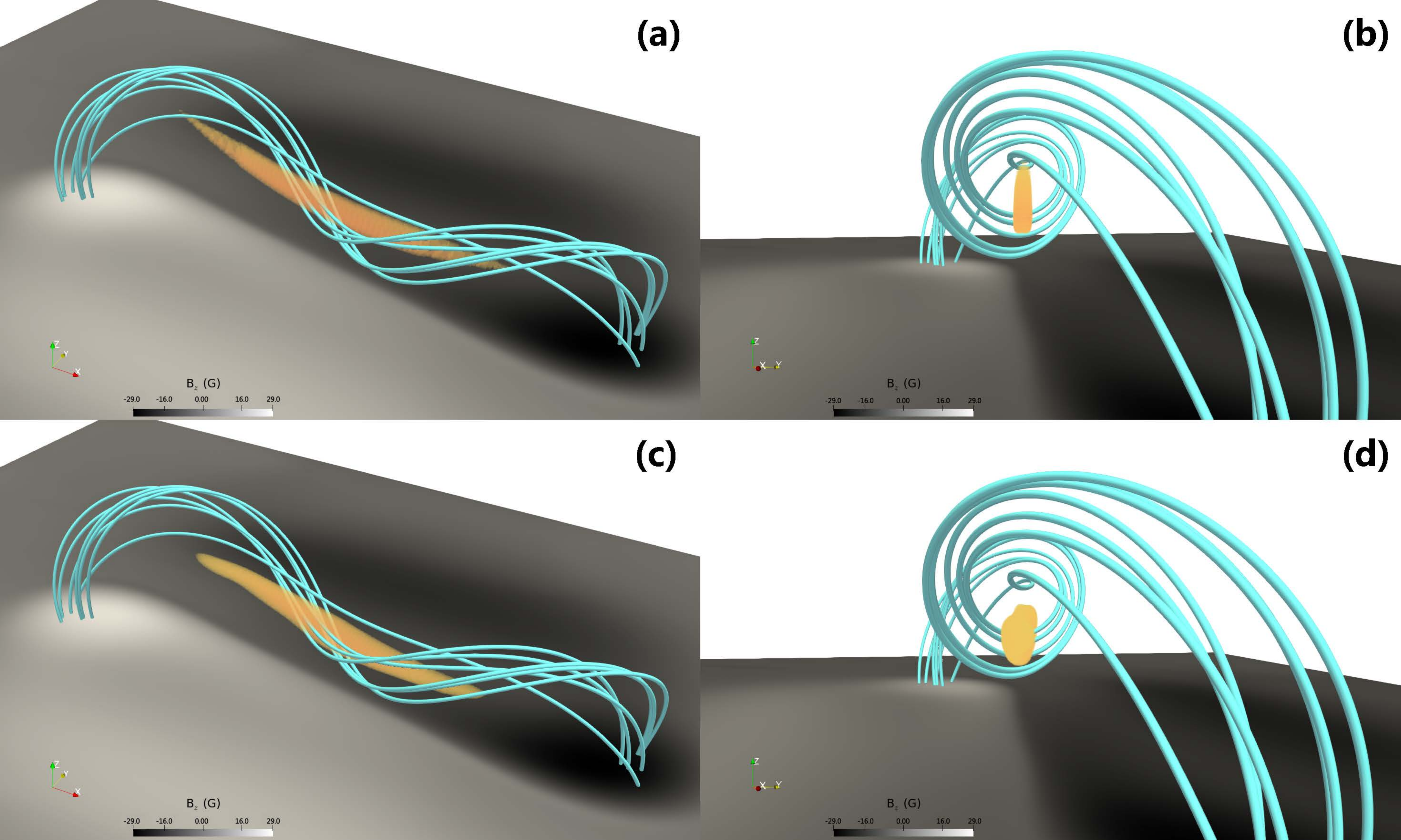}
	\caption{Panels (a, b): Two perspectives of the prominence inserted into a force-free magnetic field; Panels (c, d): Two perspectives of the prominence and the 3D magnetic field lines when the inserted prominence reaches its final equilibrium. In this figure, the yellow isosurface traces the prominence layer whose density is 20 times the background density, and the light blue lines are selected magnetic field lines enveloping the prominence. The grayscale in the bottom plane indicates the $z$-component of the magnetic field.}
    \label{fig2}
\end{figure*}

The next step is to construct a model prominence. This can be done by
performing simulations of the evaporation-condensation \citep{xia16} or
reconnection-condensation models \citep{kane17}, which are computationally
expensive. Since we do not focus on the physical process of the prominence
formation, we here choose to build a prominence in a more convenient way simply
by increasing the density by about two orders of magnitude while keeping the
gas pressure unchanged, as used by \citet{terr15} and \citet{zhou17}. Following
other authors \citep[e.g.][]{terr16} and guided by our own experience, we
choose to build the prominence by modifying the density distribution from
$\rho_{old}$ to $\rho_{new}$, which is expressed as
\begin{equation}
    {\rho _{new}}=
	\begin{cases}
        {\rho _{old}}(1 + C_{\rho}(1 + \tanh(\frac{{{l_x} - |x|}}{{{w_x}}}))(1 + \tanh(\frac{{{l_y} - |y|}}{{{w_y}}}))(1 + \tanh(\frac{{{l_z} - |z - {z_0}(x)|}}{{{w_z}}}))) & |x| < l_x; \\
        {\rho _{old}}(1 + C_{\rho}(1 + \tanh(\frac{{{l_y} - |y|}}{{{w_y}}}))(1 + \tanh(\frac{{{l_z} - |z - {z_0}(x)|}}{{{w_z}}}))) & |x| \ge l_x, \\
    \end{cases}
    \label{eq29}
\end{equation}
where $l_x=7.5$ Mm, $l_y=1.5$ Mm, $l_z=4$ Mm, $w_x=35$ Mm, $w_y=0.3$ Mm, and
$w_z = 0.8$ Mm, respectively. $C_{\rho}=50$ is a parameter used to control the
density contrast with the background corona. The parameter $z_0(x)=20+z_c-
\sqrt{\max(z_c^2-x^2, 0)}$ Mm is the initial height of the prominence
centroid, where $z_c=100$ Mm. Then, we rotate the density distribution by an
angle of $10^{\circ}$ with respect to the $z$-axis by multiplying the density
array with a rotation matrix, which makes the prominence be aligned with the
flux rope. With these operations the prominence has a maximum density 44.4
times the background one, and a temperature of $1.4\times 10^4$ K. Figures
\ref{fig2}(a) and \ref{fig2}(b) show the inserted prominence and the magnetic
field lines viewed in two different perspectives, where the yellow isosurface
traces the prominence layer whose density is 20 times the background density,
and the light blue lines represent the magnetic structure enveloping the
prominence. The approximate volume of the prominence is about 70 Mm$\times$5
Mm$\times$10 Mm with a total mass of $4.7 \times 10^{10}\unit{kg}$, which is a
typical value for a light prominence similar to previous work \citep{terr15,
terr16}. It is noted that the inserted state is not in equilibrium. Therefore,
we allow the whole system to evolve for about half an hour until the maximum
velocity within the prominence is less than 2 km s$^{-1}$, which is one-order
of magnitude smaller than the perturbation velocity used for prominence
oscillations. The relaxed state of the prominence and the field lines viewed
from two perspectives are displayed in Figures \ref{fig2}(c) and
\ref{fig2}(d), where the prominence is suspended at a height of 18.5 Mm for the
longitudinal oscillation case. We use the relaxed state as the real initial
conditions for our numerical simulations in this paper.

It is also mentioned that from this stage on, the full ideal MHD equations are
numerically solved, which means we also solve the internal energy equation
\begin{equation}
    \frac{{\partial {e_{{\mathop{\rm int}} }}}}{{\partial t}} + \nabla  \cdot ({e_{{\mathop{\rm int}} }}\bma{v}) =  - p\nabla  \cdot \bma{v},
    \label{eq27}
\end{equation}
where $e_{\mathop{\rm int}}=p/(\gamma-1)$ is the internal energy.
The heat capacity ratio $\gamma = 5 /3$ represents an adiabatic process.

\section{Perturbations and Oscillations} \label{sec3}

In order to study filament oscillations, velocity perturbations are imposed to
the prominence. Different directions of the velocity lead to longitudinal,
horizontal transverse, and vertical transverse oscillations, respectively.
Taking the longitudinal oscillations as an example, the perturbation velocity
we impose here is
\begin{equation}
  \bm{{v}_{per}}(x,y,z) = {v_1}(x,y,z)\frac{{\bm{B}(x,y,z)}}{{|B(x,y,z)|}},
  \label{eq-velo}
\end{equation}
where $v_1$ is in a form similar to Equation (\ref{eq29}), i.e.,
\begin{equation}
  {v_1}(x,y,z) = {v_0}(1 + \tanh(\frac{{{l_{vx}} - |x|}}{{{w_{vx}}}}))(1 + \tanh(\frac{{{l_{vy}} - |y|}}{{{w_{vy}}}}))(1 + \tanh(\frac{{{l_{vz}} - |z - {z_{v0}}|}}{{{w_{vz}}}})).
  \label{eq32}
\end{equation}
The parameters in Equation (\ref{eq32}) are chosen as follows so that the
perturbation region is larger than the prominence while much smaller than our
simulation box: $l_{vx}=100$ Mm, $l_{vy}=50$ Mm, $l_{vz}=4$ Mm, $w_{vx}=30$ Mm,
$w_{vy}=15$ Mm, and $w_{vz} = 10$ Mm. The height $z_{v0}=18$ Mm is a little
lower than $z_0$ in Equation (\ref{eq29}) since the height of the prominence
decreases a little after the relaxation step. $v_0$ is a constant used to
control the maximum velocity perturbation. For transverse oscillations, we just
change the direction of the velocity perturbation in Equation (\ref{eq-velo}),
making it orthogonal to the magnetic field instead.
We actually considered three methods to excite global filament
oscillations in the simulations. One is to specify the velocity perturbation in
and around the filament. The second is to add a high pressure region next to
the filament to mimic released thermal energy by nearby magnetic reconnection.
The third is to introduce a shock wave which is possibly induced by a remote
coronal mass ejection, and let the shock impact the filament.
As demonstrated by \citet{zhan13}, the oscillation characteristics are nearly the
same under impulsive high-pressure and direct velocity perturbation. Therefore, we
take the first method and include large scale perturbations only in the filament,
excluding secondary effects by an external perturbation on the filament environment. This
is numerically convenient and representative for anything that results in bulk movement of a filament.

\subsection{Longitudinal oscillations} \label{sec3.1}

\begin{figure*}
	\centering
	\includegraphics[width=\linewidth]{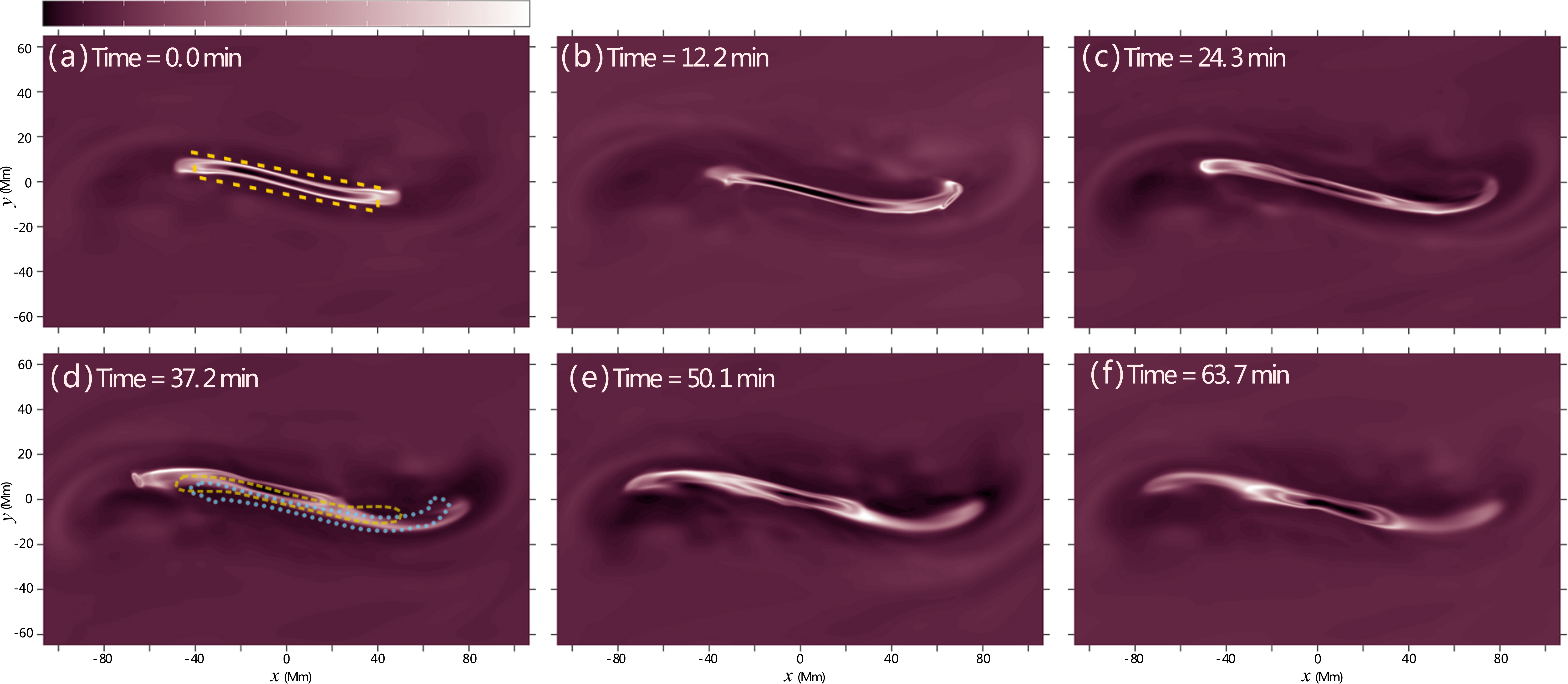}
	\caption{Top view of the synthesized EUV 211 \r{A} images of the longitudinally oscillating filament at six moments. The
   parallelogram in panel (a) marks the slice used for plotting Fig.
   \ref{fig4}, and the yellow dashed line and the cyan dotted line in panel
   (d) mark the initial and the rightmost positions of the filament.}
    \label{fig3}
\end{figure*}

To trigger a longitudinal oscillation, we simply impose a velocity perturbation
described by Equation (\ref{eq-velo}) to the filament. The velocity is aligned
with the magnetic field lines with a maximum value of $25\unit{km}
\unit{s^{-1}}$ and decreases gradually down to zero
in its neighborhood. To compare our results with observations more clearly,
synthesized emission in the extreme ultraviolet (EUV) waveband 211 \AA\ is
calculated from the simulation data. The emission in each cell of our domain is
calculated via
\begin{equation}
  {I_{\lambda}}(x,y,z) = {G_{\lambda}}(T)n_{e}^2(x,y,z),
\label{eq33}
\end{equation}
where the wavelength $\lambda$ = 211 \r{A} and $G_{\lambda}$ is the
temperature-dependent response function for the 211 \AA\ waveband, which is
obtained directly from the CHIANTI atomic database \citep{dera97,delz15}.
Figure \ref{fig3} shows a time sequence of the 211 \r{A} images of our
results from a top view. The emission is integrated along the line of sight, in
this case the $z$-direction. For simplicity, we suppose that the emissions from
the chromosphere are uniform and invariant. Thus, they are ignored in the
integral. It is seen that, as the longitudinal velocity perturbation is
exerted, the filament starts to move to the right. At $t=12.2$ minutes, it
reaches its furthest location and starts to bounce back (Figure \ref{fig3}(b)).
The filament returns to the original position at about $t=24.3$ minutes and
continues to move to the left (Figure \ref{fig3}(c)). It reaches its leftmost
position at $t=37.2$ minutes (Figure \ref{fig3}(d)). At $t=50.1$ minutes, the
filament finishes its first round of oscillation (Figure \ref{fig3}(e)) and
starts to repeat. However, as revealed by Figure \ref{fig3}(f), the amplitude
of the oscillation becomes smaller and smaller, i.e., the oscillation gradually
decays.
To see the motion more clearly, in Figure \ref{fig3}(d) we overplot the
initial boundary of the filament as the yellow dashed line whereas its
rightmost position is the cyan dotted line.
It is noticed that during the oscillation, the filament material
spreads out to form a more diffuse structure compared to the initial state.

\begin{figure*}
	\centering
	\includegraphics[width=\linewidth]{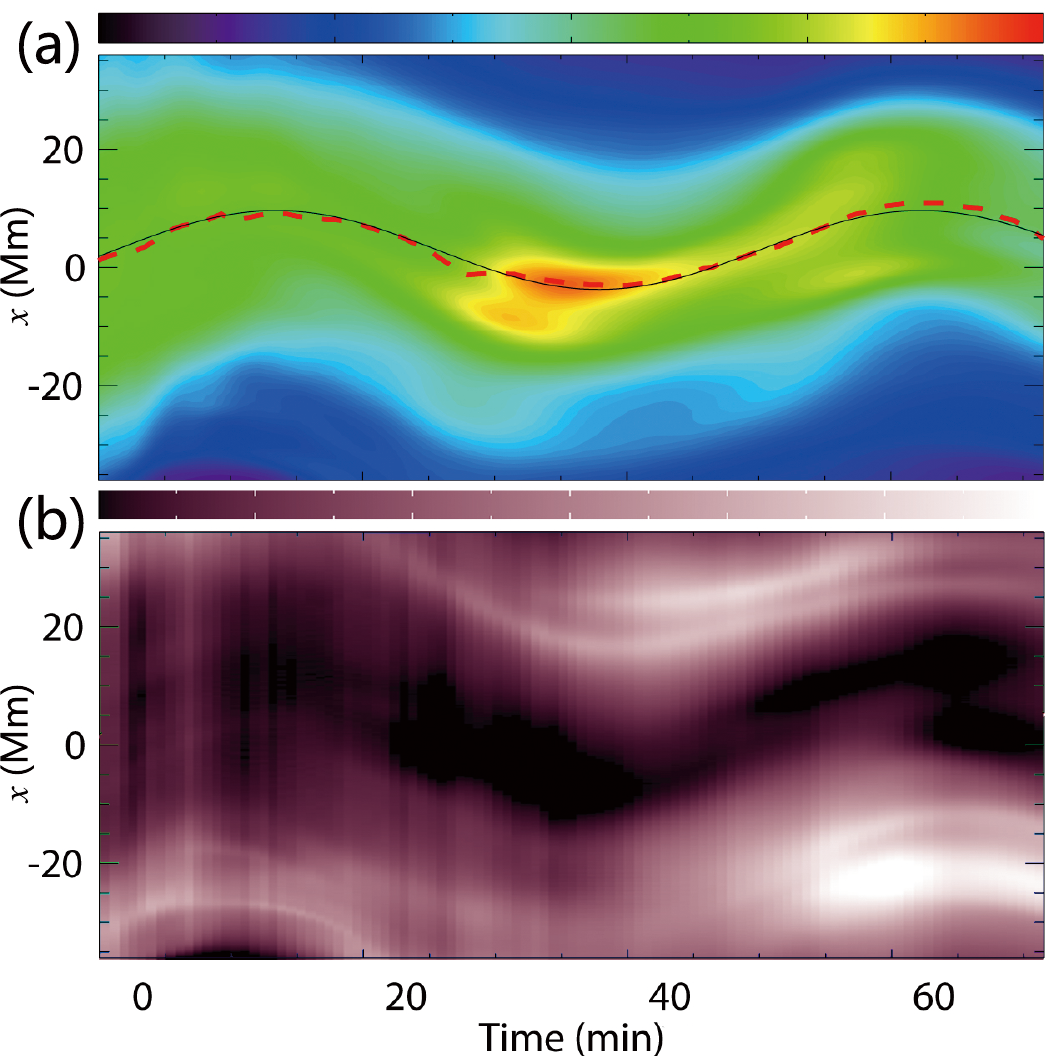}
	\caption{Time-distance diagram of the integrated density along the selected axis with a width of 5 Mm (see Figure~\ref{fig3}(a)). Red dashed line indicates the position of filament centroid, whereas the black solid line is the fitting result based on a decayed sine function.Panel (b): The synthesized EUV 211 \r{A}
     image of panel (a).}

    \label{fig4}
\end{figure*}

In order to display the longitudinal oscillation more clearly, we trace the
density distribution along the main axis of the filament. The axis is taken to
be parallel to the $x$-$y$ plane at $z=18.5$ Mm, and is skewed from the
$x$-axis by $10^{\circ}$ in order to be aligned with the filament.
Since the motion is not exactly along this selected axis, the axis has a width of 5 Mm in the $y$-direction, as marked by the yellow parallelogram in Figure \ref{fig3}(a).
The evolution of the integrated density distribution along the main axis is plotted in the
time-distance diagram in Figure \ref{fig4}. It reveals that the filament
experiences a decayed oscillation. We further calculate the centroids of the
dense plasma along the main axis at individual times, which are represented by
the red dashed line in Figure \ref{fig4}. The positions of these centroids are then
fit with a decayed sinusoidal function $d=d_0 \mathrm{e}^{-t/\tau} \sin (2\pi
t/P+\phi)$, where $d$ is the displacement of the filament, $d_0$ is the
amplitude, $P$ is the oscillation period, $\tau$ is the decay time, and
$\phi$ is the initial phase angle. The fitting results in an oscillation
period of $P=48.8$ minutes and a decay time $\tau=86.5$ minutes. The fitted
profile is overplotted on Figure \ref{fig4} as the black solid line.
The corresponding 211 \r{A} image is plotted in Figure \ref{fig4}(b)
for comparison.

\subsection{Horizontal transverse oscillations} \label{sec3.2}

By changing the perturbation velocity from Equation (\ref{eq-velo}) to $v_x=
v_1 \rm{min}(B_y, 0)/|\bm{B}|$ and $v_y=v_1B_x/|\bm{B}|$, we can excite the
horizontal transverse oscillation of the filament.

\begin{figure*}
	\centering
	\includegraphics[width=\linewidth]{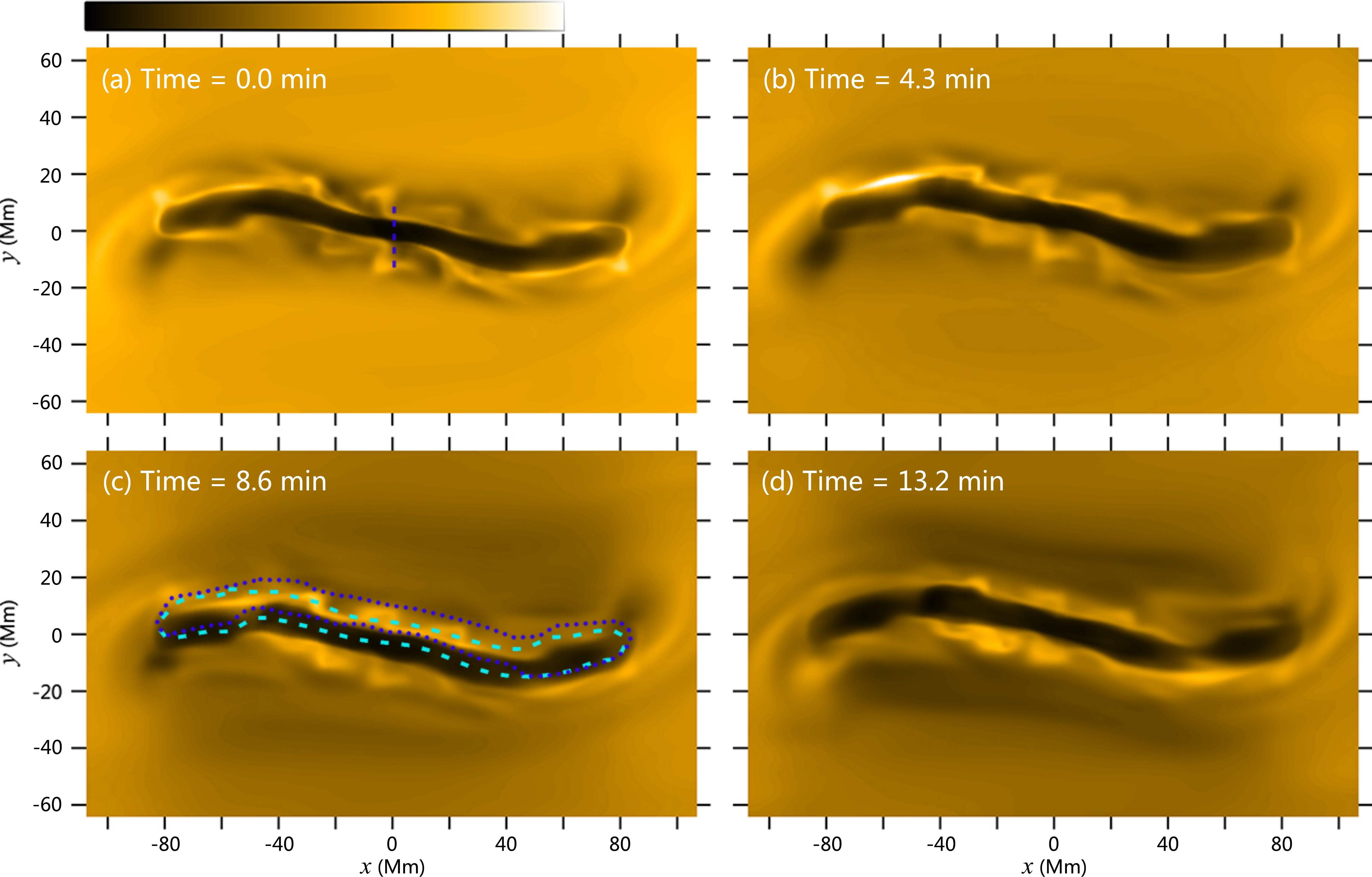}
	\caption{Top view of the synthesized EUV 171 \r{A} images at four moments in the case where the filament is experiencing horizontal transverse oscillations. The cyan dashed line and the blue dotted
     line in panel (c) indicate the initial and the uppermost positions of the
     prominence, respectively.}
    \label{fig5}
\end{figure*}

The evolution of the synthesized EUV 171 \AA\ images viewed from the top is
displayed in Figure \ref{fig5}. Similarly, the emission from the chromosphere
is not included in the calculation of the EUV intensity. From the figure, it
is seen that at $t=3.2$ minutes, the filament reaches its furthest position
in the positive $y$-direction (Figure \ref{fig5}(b)), and starts to return to the
original location. At $t=8.2$ minutes, the filament moves to its furthest
position in the negative $y$-direction, as indicated by Figure \ref{fig5}(c).
The filament returns to its equilibrium position at $t=10.7$ minutes. After
that, it repeats its oscillation, but with a smaller amplitude, as revealed by
Figure \ref{fig5}(d).
Similarly to the longitudinal oscillation, the
initial and the uppermost positions of the prominence boundary are respectively
indicated by the cyan dashed line and the blue dotted line in Figure
\ref{fig5}(c).

\begin{figure*}
	\centering
	\includegraphics[width=\linewidth]{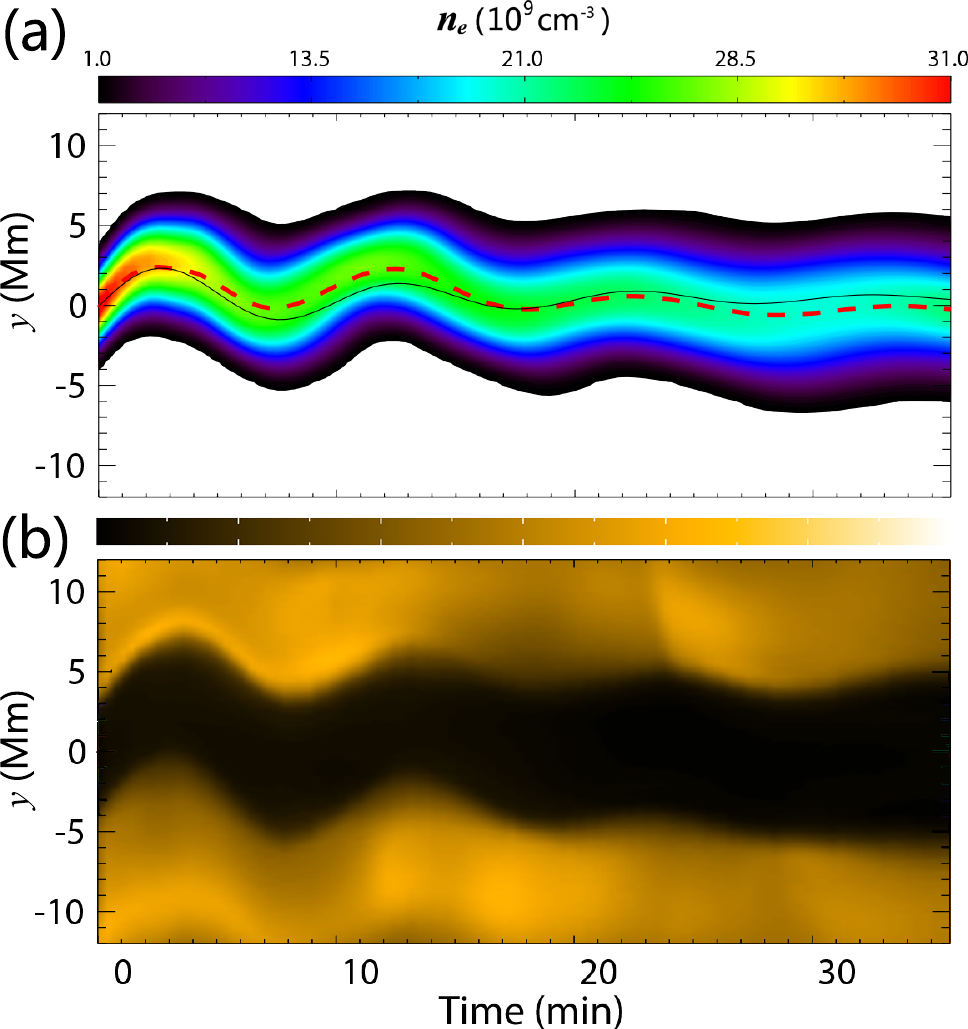}
	\caption{Time-distance diagram of the density along the $y$-axis (taken along the blue dashed line in Fig. \ref{fig5}(a)).
The red dashed line indicates the position of filament centroid, whereas the black solid line is the fitting result based on a decayed sine function. Panel (b): The synthesized EUV 171 \r{A} image of panel (a).}
    \label{fig6}
\end{figure*}

In order to show the lateral displacement more clearly, we take a slice across
the filament in the $y$-direction at $z$ = 18.0 Mm,
as indicated by the cyan dashed line in Figure \ref{fig5}(a).
The evolution of density
along the slice is displayed in the time-distance diagram in Figure \ref{fig6},
where the red dashed line describes the evolution of the prominence centroid.
Its displacement is fit with a decayed sinusoidal
function $d=d_0 \mathrm{e}^{-t/\tau} \sin (2\pi t/P+\phi)$, which leads to a
period of 10.1 minutes, and a decay time $\tau=17.5$ minutes. The fitting is shown by the black solid line.
The corresponding 171 \r{A} image is plotted in Figure \ref{fig6}(b)
for comparison.

\subsection{Vertical transverse oscillations} \label{sec3.3}

Once the perturbation velocity in Equation \ref{eq-velo} is modified to
$\bm{{v}}_{per}(x, y, z) = {v_1}(x,y,z)\hat{\mathbf{{e}}}_z$, we can excite
vertical transverse oscillations. Although the velocity direction is not
exactly perpendicular to the magnetic field lines, the deviation is minor
since the magnetic field is nearly horizontal inside the prominence.

\begin{figure*}
	\centering
	\includegraphics[width=\linewidth]{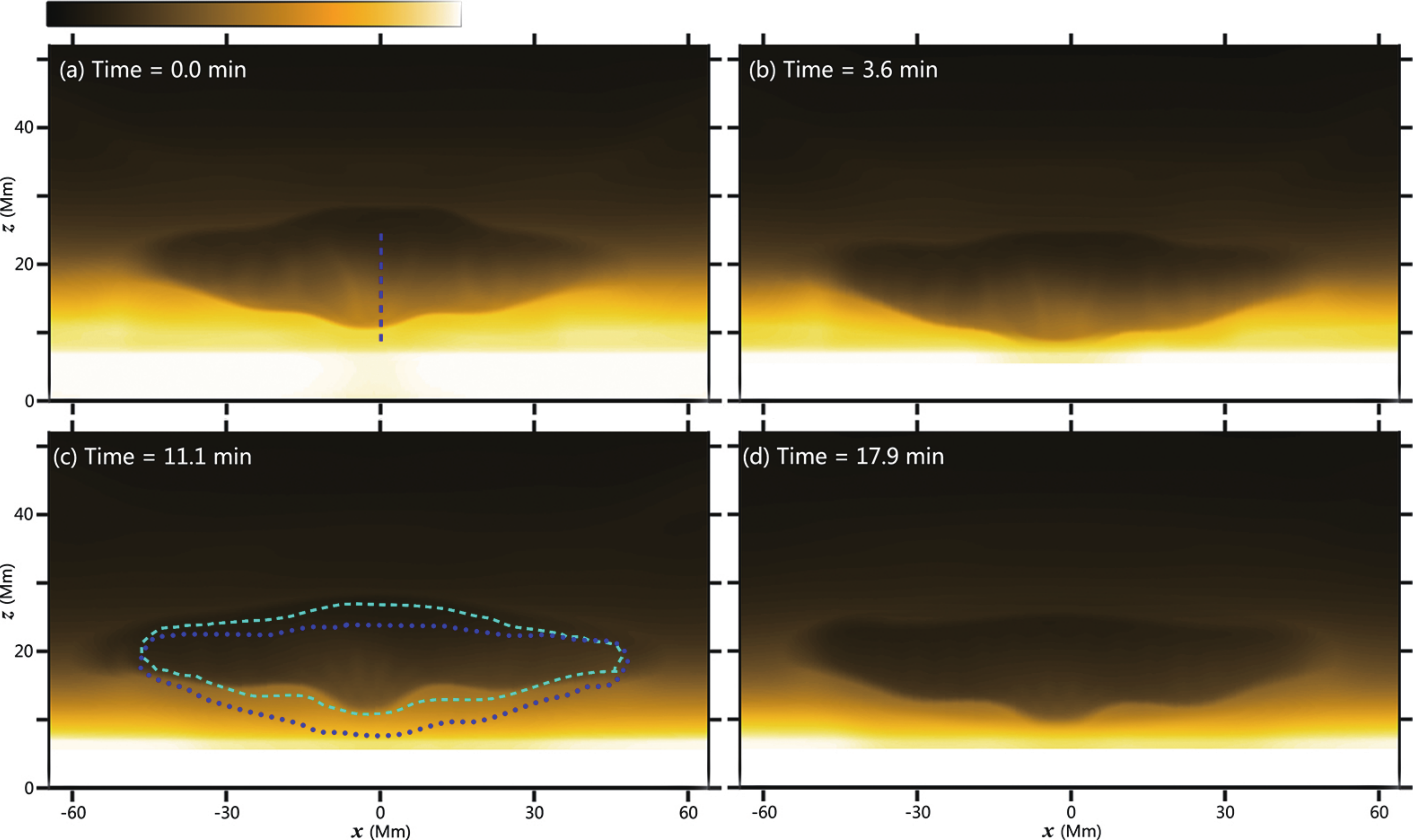}
	\caption{Side view of the synthesized EUV 171 \r{A} images at four moments in the case where the filament is experiencing vertical transverse oscillations. The cyan dashed line and the blue dotted
     line in panel (c) indicate the initial and the extremal positions of the
     prominence, respectively.}
    \label{fig7}
\end{figure*}

Again, synthesized 171 \AA\ images are used to show the dynamics of the
filament viewed from the side, i.e., the $y$-direction. The results are
displayed in Figure \ref{fig7}, where the chromosphere is colored in
white since it does not change too much, and its features would distract
the attention of the readers. It is seen that at $t=3.6$ minutes, as revealed
by Figure \ref{fig7}(b), the prominence goes down to its lowest height
(Figure \ref{fig7}(c)) and then starts to bounce back. At $t=11.1$ minutes, the
prominence reaches its highest position. The prominence reaches its minimum
height again at $t=17.9$ minutes, as shown in Figure \ref{fig7}(d).
Again, the initial and the lowest positions of the prominence are
respectively indicated by the cyan dashed line and the navy blue dotted line in
Figure \ref{fig7}(c). Comparing
panels (d) and (b), we can see that the oscillation amplitude is decaying.

\begin{figure*}
	\centering
	\includegraphics[width=\linewidth]{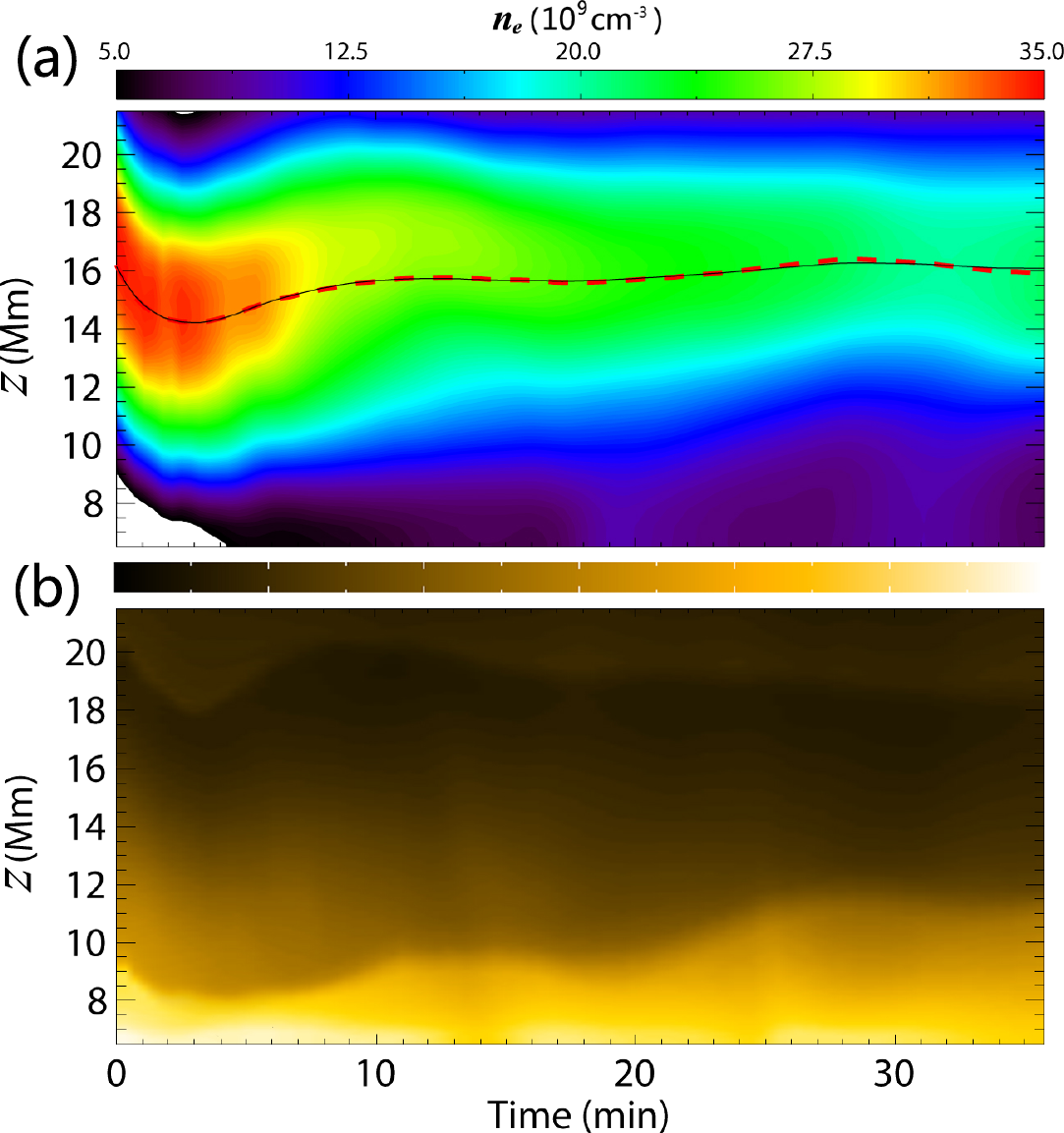}
	\caption{Time-distance diagram of the density along the $z$-axis. The red dashed line indicates the position of prominence centroid, whereas the black solid line is the fitting result based on a decayed sine function. Panel (b): The synthesized EUV 171 \r{A} image of panel (a). }
    \label{fig8}
\end{figure*}

In order to reveal the vertical oscillation more quantitatively, we examine
the density distribution along the $z$-axis, which crosses the prominence
center.
The slice is marked by the cyan dashed line in Figure \ref{fig7}(a).
The evolution of the density distribution along the $z$-axis is
displayed in the time-distance diagram in Figure \ref{fig8}, from which the
decayed oscillation is evident. The mass center of the prominence is
represented by the dashed line, and the position evolution is fit with a
decayed sine function $z=z_0+A_0 \mathrm{e}^{-t/\tau} \sin (2\pi t/P+\phi)$,
where $z_0$ is the initial height, $A_0$ is the initial amplitude of the
oscillation, $P$ is the period, $\tau$ is the decay time, and $\phi$ is the
initial phase angle. The resulting period is 14.0 minutes.
The corresponding 171 \r{A} image is plotted in Figure \ref{fig8}(b)
for comparison.

\section{Discussions} \label{sec4}

Prominence oscillations are a very interesting phenomenon. They can not only
be applied as a potential precursor for coronal mass ejections \citep{chen08,
pare14, mash16, zhou16}, but also can be used to diagnose the coronal magnetic
field \citep{arre12}. As a part of coronal seismology  \citep{naka05, andr09},
prominence seismology seems more complicated than its coronal loop
counterpart due to the complex structure of the former. Among all the
parameters obtained from observations, oscillation periods and damping time
are two straightforward quantities that can be used to constrain the restoring
force and the damping mechanisms. Many linear models have been established
for different restoring forces \citep{oliv02} and damping mechanisms
\citep{oliv09}, and the validity of these linear models should be verified by
nonlinear MHD simulations. In this paper, we performed 3D MHD simulations
of prominence oscillations, and concentrated on the restoring forces only,
leaving the damping mechanism for future work.
Investigating the damping mechanisms requires a much higher spatial
resolution in the numerical simulations for the physical processes to stand
out from the numerical dissipation.

To obtain a model prominence embedded in a magnetic flux rope, we first created
a nearly force-free flux rope via evolving the bottom boundary conditions in
an isothermal MHD simulation. Then, the density was increased and the
temperature was
decreased inside an ellipsoidal volume. Such distributions, which are not in
mechanical equilibrium, gradually evolved to an equilibrium state through
relaxation. We must be aware of some artifacts of this model prominence due to
our ideal MHD assumptions. For example, looking at Equation (\ref{eq29}), we
may find that we set a fairly large PCTR in the spine direction, which may not
be true in observations. Then we imposed an impulsive perturbation over the
prominence, which could be due to a passing EUV wave \citep{shen14}. The
direction of the perturbation was controlled in order to excite longitudinal,
horizontal transverse, and vertical transverse oscillations, respectively.

\subsection{Pendulum model for the Longitudinal oscillations}

As shown by Figure \ref{fig4}, our numerical results indicate that the centroid
of the prominence oscillates with a period of about 48.8 minutes, which is in the
typical range for the observed longitudinal oscillations \citep{trip09}. In
order to check the pendulum model, we extract the magnetic field line across
the prominence centroid, and calculate the curvature radius near the magnetic
dip, which is found to be $R=$ 52.6 Mm. According to the pendulum model, the
theoretical period of the longitudinal oscillation should be $P=2\pi
\sqrt{R/g}$ = 45.9 minutes, which is very close to the oscillation period in our
3D MHD simulations. Such consistency confirms that the field-aligned component
of gravity is responsible for the restoring force for filament longitudinal
oscillations.

\begin{figure*}
	\centering
	\includegraphics[width=\linewidth]{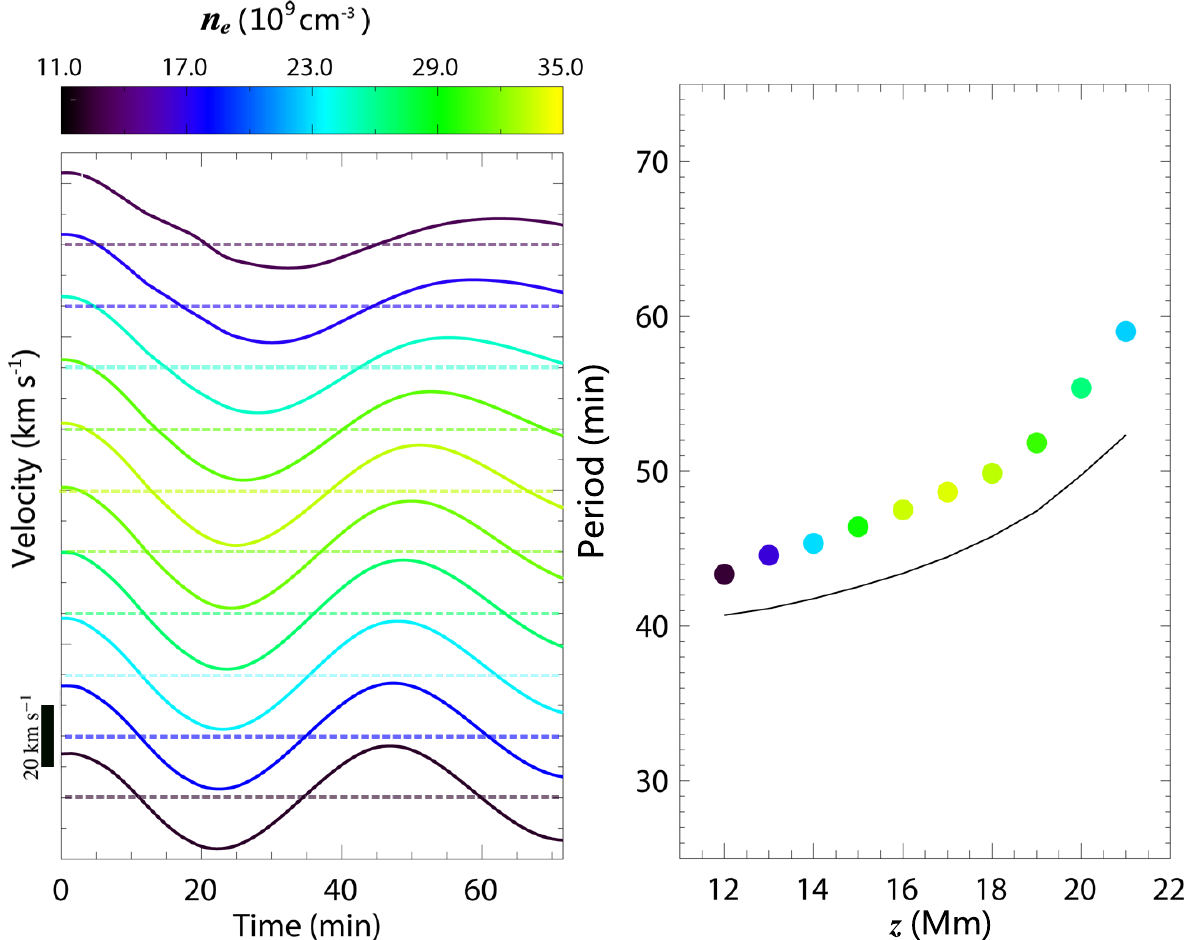}
	\caption{Left panel: Evolution of the density-weighted average field-aligned velocity of 10 selected magnetic field lines in the longitudinal oscillation case. The horizontal axis is physical time and the velocity profiles of different field lines are stacked one by one in a sequence of their heights. The scale for the velocity is plotted at the lower-left corner. Different colors indicate different initial densities at the center of the magnetic dips of different field lines, and the color scale for the density is shown at the top of this panel. Right panel: Oscillation periods of the 10 selected magnetic field lines at different heights, where the solid circles are derived from our simulations, and the black solid line represents the theoretical values calculated from the pendulum model. The color of each circle has the same meaning as panel (a), indicating the averaged density along the field line.}
    \label{fig9}
\end{figure*}

On the other hand, as mentioned by \cite{terr13}, different parts of the
prominence at different heights may not oscillate in phase. They oscillate with
different periods. In order to check it, ten field lines going through the
$z$-axis at $t$ = 0 are selected, from heights between $z$ = 12 Mm to $z$ = 21 Mm,
with a separation of 1 Mm.  Following the analysis in \citet{luna16}, the
density-weighted average field-aligned velocity is calculated as
\begin{equation}
  {{{v_{\parallel i}}(t) = \int {{v_\parallel }} ({s_i},t)\rho ({s_i},t)d{s_i}} \mathord{\left/
 {\vphantom {{{v_{\parallel \, i}}(t) = \int {{v_\parallel }} ({s_i},t)\rho ({s_i},t)d{s_i}} {\int {\rho ({s_i},t)d{s_i}} }}} \right.
 \kern-\nulldelimiterspace} {\int {\rho ({s_i},t)d{s_i}} }},
  \label{eq41}
\end{equation}
where $i$ means the $i$-th field line we select and $s_i$ is the 1D arc length
along the field line. Here, the velocity, instead of the displacement, is
considered because the magnetic lines themselves are also moving. The results
are plotted in Figure \ref{fig9}(a), where the $x$-axis is the physical time in
minutes. Ten profiles with different colors are stacked one by one in the
sequence of
height, and the zero velocity for each profile is indicated by the dashed line with
the same color as the corresponding velocity profile. Each profile is fit with
a decayed sine function, and the resulting periods, as a function of height of
the magnetic dip, are plotted as solid circles in Figure \ref{fig9}(b). It is
seen that the oscillation period increases from 43 minutes to 59 minutes as the
height of the plasma increases.  According to the pendulum model, such a result
means that the curvature radius of the magnetic dips becomes larger and larger
at higher positions. To confirm it, we calculate the curvature radius ($R$) for
each magnetic field line at the dip site in our numerical results. It is noted
that the magnetic dip is not a perfect circle. Therefore, we take an average
value within $\pm10$ Mm near the center of the magnetic dip, assuming that the
magnetic field lines do not deform during the oscillations. Indeed it is found
that the curvature radius of the magnetic dips increases with height. Based on
the pendulum model, i.e., $P=2\pi\sqrt{R/g}$, the theoretical periods of the
longitudinal oscillations along these field lines are calculated and plotted
as the solid line in Figure \ref{fig9}(b).  We can see that the theoretical
results are roughly in agreement with the numerical results, further confirming
that the field-aligned component of gravity serves as the restoring force for
filament longitudinal oscillations. However, it is noticed in Figure
\ref{fig9}(b) that the theoretical results are systematically smaller than the
3D numerical results. The deviation is about 10\%.

Two conditions are required for the pendulum model to work well: (1) The
curvature radius of the magnetic dip should not be too large; and (2) The
magnetic field line does not deform significantly during prominence
oscillations. For the first
requirement, \citet{luna12a} introduced a reference radius of curvature,
$R_{lim}$ in their Equation (33). When the curvature radius of a magnetic dip,
$R$, is smaller than $R_{lim}$, the field-aligned component of gravity
overwhelms the gas pressure gradient, and the gas pressure can be neglected.
In our simulation, $R_{lim}$ is about 450 Mm, and the curvature radius of the
magnetic dips is $\sim$50 Mm, several times smaller than the reference radius
$R_{lim}$. In our previous 1D simulations \citep{zhan13, zhou17}, the curvature
radius of the magnetic dips was also smaller than $R_{lim}$, which is why both
the 3D simulations in this paper and our previous simulations showed
consistency with the pendulum model.  Regarding the second requirement, it is
often argued that the plasma $\beta$ should be much smaller than unity.
Observations indicate that the magnetic field strength of a quiescent
prominence is generally 10--30 G \citep{bomm94, mere06}. The corresponding
plasma $\beta=0.05 {n \over {10^{11}~\rm{cm}^{-3}}} {T \over {10^4~\rm{K}}}
({B \over {10~\rm{G}}})^{-2}$ would be much smaller than unity for the typical
density and temperature. In our simulation case, the plasma $\beta$ inside the
prominence is $\sim$0.01, much smaller than unity. Therefore, it seems that
the two requirements are both satisfied.

While the simulation results are quite consistent with the pendulum model, it
is still worthwhile to mention that the actual period of the longitudinal
oscillations in our 3D simulations is systematically larger than that predicted
by the pendulum model. This feature cannot be explained by the additional
effect of gas pressure, as the inclusion of extra gas pressure gradient would
increase the restoring force, hence shorten the oscillation period. For
example, \citet{luna12a} considered the combination of gravity and gas pressure.
They found that the resulting oscillation period is smaller than that
determined by gravity only. The possible reason for the larger period in
simulations is the deformation of the magnetic field line \citep{li12}, which
changes the local curvature radius dynamically. Whether the magnetic field can
be deformed is not determined by the plasma $\beta$ only. We think that another
parameter should be the ratio of the gravity to the magnetic pressure
\begin{equation}
\delta={{\rho gL}\over {B^2/2\mu_0}}=11.5{n \over {10^{11}~\rm{cm}^{-3}}} {L \over {100~\rm{Mm}}} ({B \over {10~\rm{G}}})^{-2},
\label{eq42}
\end{equation}
where $n$ is the number density of the prominence, $L$ is the length of the
prominence thread, and $B$ is the magnetic field. For the typical values of
these parameters in our simulation, the newly defined dimensionless parameter
$\delta$ is round unity, i.e., the gravity is comparable with the magnetic
pressure force. Therefore, the deformation of the magnetic field lines is not
negligible. The gravity-induced deformation makes the magnetic dip flatter,
which increases the oscillation period.

\subsection{Explanation for the Horizontal Transverse Oscillation}

Our numerical results show that the horizontal transverse oscillation has a
period of $9.9\pm0.4$ minutes. If we use the simple 1D string model, the
oscillation period defined by Equation (\ref{eq11}) \citep{joar92} is about
17 minutes, which is over 60\% larger than the actual period. The reason for the
discrepancy is that the prominence was assumed to be infinitely wide in their
1D model. While considering the finite width of prominences, \citet{diaz01}
improved the 2D model of \citet{joar97}, and derived the new dispersion
relation. Even for the fundamental mode of this model, the new equations become
transcendental so that no analytic solution can be given, and the equations
have to be numerically solved. The most important parameter required for this
model, other than those needed for Equation~(\ref{eq11}), is the thickness of
the prominence in the transverse direction, which can also be obtained directly
from our simulation. Other parameters needed for this model, such as the
density contrast, can also be obtained through averaging, though this model is
not very sensitive to them. It is noted that the Alfv\'en speed used in this
model is not taken inside the prominence, but outside it. Therefore, we take an
average value of the Alfv\'en speed along the magnetic line in the corona,
excluding the prominence part. The resulting oscillation period is $\sim$10
minutes, which is very close to our numerical simulation.

\begin{figure*}
	\centering
	\includegraphics[width=\linewidth]{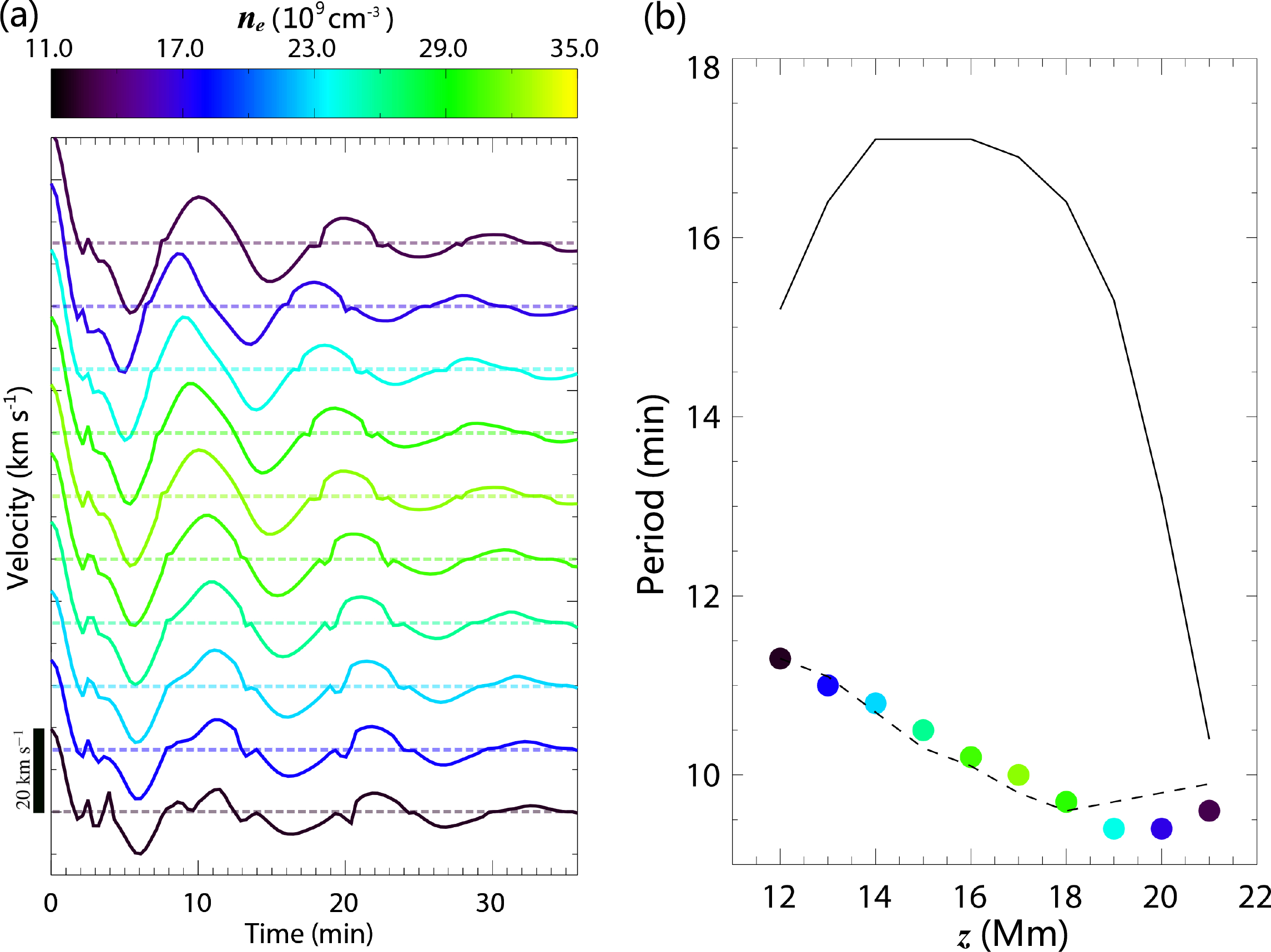}
	\caption{Similar to Figure \ref{fig9}, but for the $y$-component of velocity in the horizontal transverse oscillation case. Two models are compared to the simulations at right: a 2D slab model matches best.}
    \label{fig10}
\end{figure*}

Similar to the longitudinal oscillation case, we also pick 10 different
magnetic lines that go through the $z$-axis at $t = 0$, and plot the time
evolution of their density-weighted horizontal transverse velocity in Figure
\ref{fig10}(a), where the zero velocity for each velocity profile is indicated
by the dashed line with the corresponding color. Compared to the longitudinal
oscillation, the horizontal transverse oscillation presents a much shorter
period, which is in accordance with observations \citep{trip09}. Their
oscillation periods can also be obtained by fitting the velocity profiles with
decayed sine functions. However, since there are more fluctuations in the
velocity profiles, we perform wavelet spectral analysis on the velocity evolutions, and
the resulting period as a function of the height of the magnetic dip is
displayed as solid circles in Figure \ref{fig10}(b). It is seen that, opposite
to the longitudinal case, the oscillation period decreases slightly with height.

The above-mentioned two linear models are compared with our simulation results.
For the 1D string model, with all the parameters required for Equation
(\ref{eq11}) extracted from the simulations, the resulting periods for the
ten magnetic field lines are plotted as the solid line in Figure \ref{fig10}(b).
It is seen that the theoretical periods deviate from the 3D simulation results
significantly. For the 2D slab model \citep{diaz01}, with all the parameters in
each magnetic field line included, the calculated oscillation periods are
overplotted in Figure \ref{fig10}(b) as the dashed line. We can see that the
2D slab model matches with the 3D simulation remarkedly well.

It should be mentioned that the horizontal transverse oscillation periods of
different parts of the prominence are not much different. It seems that the
prominence oscillates horizontally as a whole.

\subsection{Explanation for the Vertical Transverse Oscillation}

The numerical results indicate that the vertical transverse oscillation of the
prominence centroid has a period of $14.1\pm1.5$ minutes. If we use the simple
1D string model, the oscillation period defined by Equation (\ref{eq11})
\citep{joar92} is about 17 minutes, which is $\sim$20\% larger than the actual
period, which seems not so bad. It is noted here that the vertical and
horizontal transverse oscillations are not distinguishable in the 1D string
model. If we use the 2D slab model \citet{diaz01}, however, the calculated
oscillation period is 14 minutes, which is almost the same as our 3D
numerical simulations.

\begin{figure*}
	\centering
	\includegraphics[width=\linewidth]{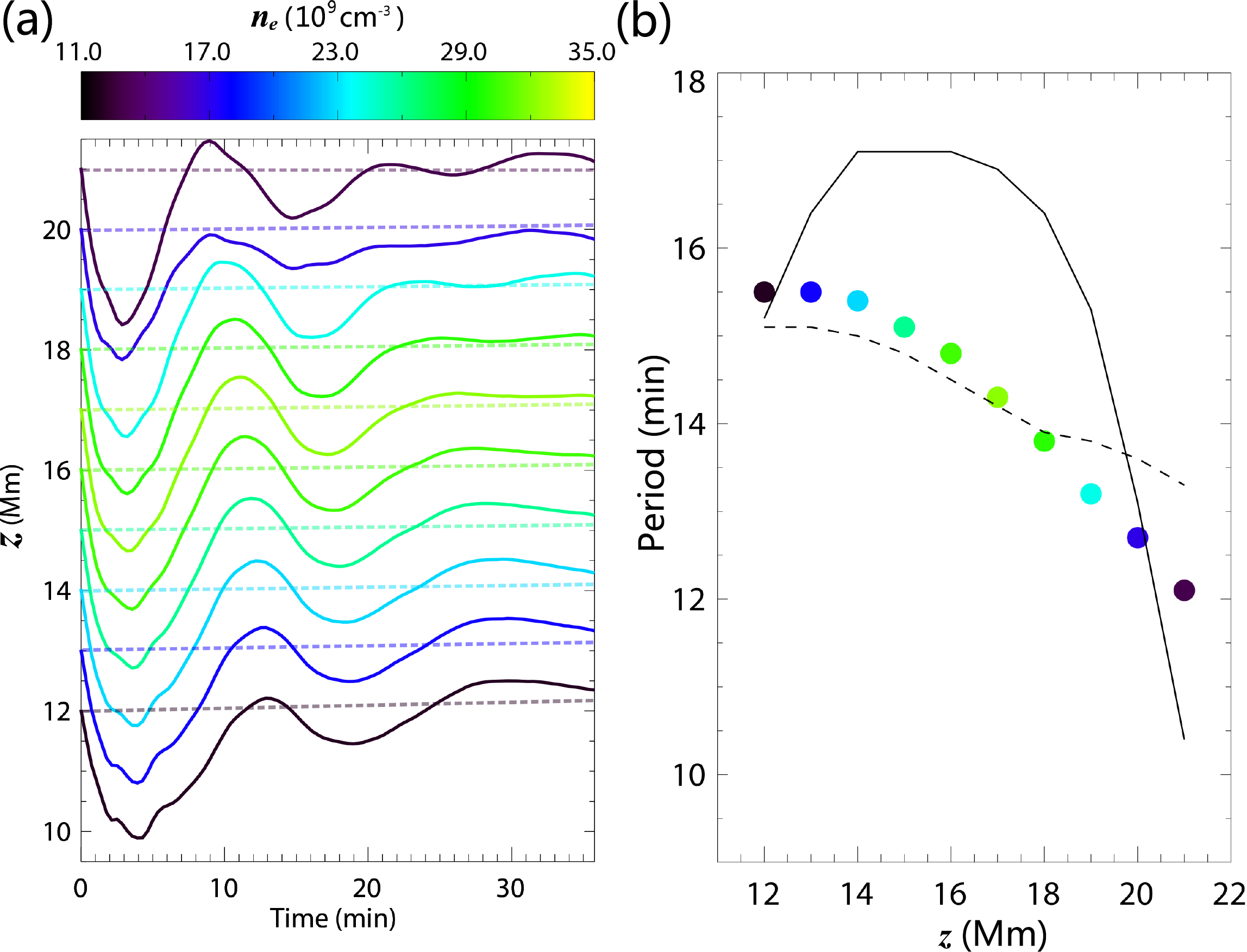}
	\caption{Similar to Figure \ref{fig9}, but for the displacement of the magnetic field lines in the $z$-direction in the vertical transverse oscillation case.}
    \label{fig11}
\end{figure*}

In order to check whether different parts of the prominence oscillate
synchronously in the vertical transverse case, we select the same 10 magnetic
lines as before to analyze their motions in detail.  Different from the previous
two subsections, for the vertical transverse oscillation we simply plot the
displacements of these field lines in the $z$-direction because the magnetic
lines near the prominence are almost horizontal, and the vertical displacement
reflects the motion directly. The time evolutions of their displacements are
plotted in Figure \ref{fig11}(a) as solid lines with different colors, where
the equilibrium location for each line is indicated by the dashes with the
corresponding color. We can see that the vertical motions in Figure
\ref{fig11}(a) are much smoother than those for the horizontal transverse ones
in Figure \ref{fig10}(a). Therefore, it is easy to fit these lines with decayed
sine functions. The resulting period as a function of the height of the
magnetic dip is displayed as solid circles in Figure \ref{fig11}(b). It is seen
that the oscillation period decreases with height, with the same tendency as
the horizontal transverse oscillation, but more drastically. Actually the
difference of the oscillation periods among the ten field lines is evident even
by directly looking at the velocity profiles in Figure \ref{fig11}(a). It seems
that the prominence is not oscillating collectively.

The two linear models are compared with our simulation results in this case as
well. For the 1D string model, with all the parameters of each field line
required for Equation (\ref{eq11}) extracted from the simulations, the
resulting periods for the ten magnetic field lines are plotted as the solid
line in Figure \ref{fig11}(b). It is seen that the theoretical periods deviate
from the 3D simulation results significantly for most field lines. For the 2D
slab model \citep{diaz01}, with all the parameters in each magnetic field line
included (note that the vertical thickness of the prominence is the same for
all the field lines), the calculated oscillation periods are overplotted in
Figure \ref{fig11}(b) as the dashed line. We can see that the 2D slab model
better matches with the 3D simulation.

It is noted that, as seen from Figure \ref{fig11}(a), all the magnetic field
lines have a tendency to rise slightly in altitude during oscillation. This is
because some of the cold prominence material (about 20\% of the total mass)
drains down to the solar surface while oscillating vertically. The reduced
gravity leads to the slow rise of the prominence. Another prominent feature
of the vertical transverse oscillation, compared to the horizontal transverse
oscillation, is that the two modes have different oscillation periods. The
satisfactory matching between the 2D slab model and the 3D simulations implies
that the difference of the oscillation period is simply due to the different
horizontal width from the vertical thickness of the prominence.
In our case, the vertical thickness of the prominence is about 3 times larger
than the horizontal width. In this sense, it seems that the effect of the
curvature of the 3D magnetic field lines is negligible. Such a characteristic
was also valid for coronal loop oscillations \citep{vand09, terr16}. It is also
inferred that when the horizontal and vertical transverse oscillations are
observed to have similar periods \citep[e.g.,][]{ning09}, it probably implies
that the aspect ratio of the cross section of the prominence is close to
unity.

\subsection{Restoring Force for the Transverse Oscillations}

\begin{figure*}
	\centering
	\includegraphics[width=\linewidth]{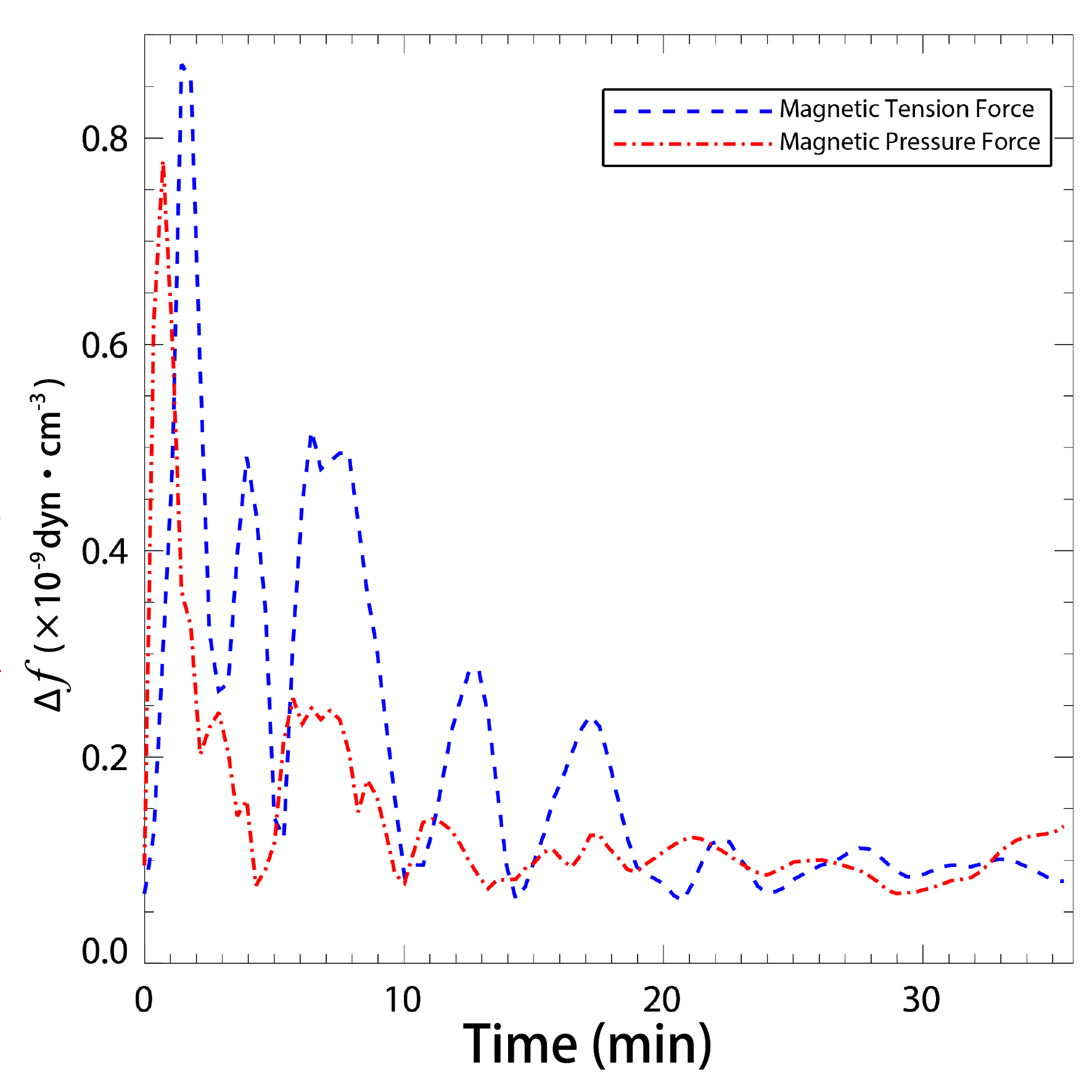}
	\caption{Time evolution of the normalized magnetic tension force and magnetic pressure force averaged along a selected magnetic field line. It shows that magnetic tension dominates in the restoring force during most of the oscillation. }
    \label{fig12}
\end{figure*}

It is conceivable that the restoring force for the transverse oscillations is
the Lorentz force, i.e., $\bma{J} \times \bma{B}$, which can be decomposed into
the magnetic pressure force and the magnetic tension force as follows:
\begin{equation}
  \bma{J} \times \bma{B}=-\nabla ({\bm{B}^2}/2{\mu_0})+\frac{{\bma{B}\cdot
	\nabla \bma{B}}}{{{\mu_0}}},
  \label{eq43}
\end{equation}
where the first term on the right-hand side is the magnetic pressure force and
the second term is the magnetic tension force. To investigate which term is
dominant in the case of the horizontal transverse oscillation, we select a
magnetic line which goes through the $z$-axis at $z=17$ Mm when $t$=0, and
calculate the horizontal components of the Lorentz force, the tension force,
and the magnetic pressure force along this field line. Then, we define the
averaged change of the magnetic tension force and the magnetic pressure force
weighted by the deviation of the Lorentz force from the initial state as
follows:
\begin{equation}
  \Delta {f}(t) = \frac{\int {|{f}(t)-{f(0)}||{f_L}(t)-{f_{L}(0)}|ds}}
	{\int {|{f_L}(t) - {f_{L}(0)}|ds}},
  \label{eq44}
\end{equation}
where $f$ stands for the magnetic tension force or the magnetic pressure force,
and $f_L$ stands for the Lorentz force. The time evolution of $\Delta f$ is
plotted in Figure \ref{fig12}, where the blue line corresponds to the magnetic
tension force and the red line represents the magnetic pressure force. It is
revealed that the magnetic pressure force overwhelms for the first one minute
only, and the magnetic tension force becomes dominant since then.
It is interesting to notice the periodic variations of both the
normalized unsigned magnetic tension and magnetic pressure force. For the
dominant tension force, it shows a period of $\sim$5 minutes (except the second
peak), which is exactly half the filament oscillation period as expected. For
the subordinate magnetic pressure force, it shows higher frequency fluctuations
in addition to the $\sim$5-minute oscillation. The higher-frequency
oscillations might be due to other oscillation modes, such as the sausage mode.
Although the magnetic pressure force is not important for the consideration of
the restoring force in this paper, its multi-period oscillations definitely
deserve further investigations.
We also did the same analysis for the vertical transverse oscillations, and
found that the result is similar: Only during the first one minute, the
magnetic pressure is dominant. Since then, the magnetic tension force is
always dominant during oscillations. Therefore, it is reasonable to assume the
magnetic tension force as the restoring force for the period analysis as used
widely in literature \citep[see][for a review]{arre12}.

To summarize, we performed 3D MHD numerical simulations of prominence
oscillations, including the longitudinal one and the transverse ones (both
horizontal and vertical), with the purpose to compare its oscillation periods
with various models and examine their restoring forces. It is confirmed that
the magnetic field-aligned component of gravity is responsible for longitudinal
oscillations, and magnetic tension force is the main restoring force for
transverse oscillation. Whereas the oscillation period of the longitudinal
oscillation can be determined by the pendulum model, with an error up to 20\%
for the shallowest dips present in our modeled flux rope, the period of the
transverse oscillation can be nicely determined by the 2D slab model described
by \citet{diaz01}, where the width (or thickness) of the prominence in the
oscillation direction is also an important parameter.

It should be noted here that the model prominence in our simulation is a
monolithic body. However, prominences are observed to be composed of many thin
threads \citep{lin10}, and these threads might oscillate with the same period
\citep{lin04} or different periods \citep{mash09b, mash09a}. The thread-thread
interactions have been investigated in 1D \citep{zhou17} by simulations and in
2D via linear analysis \citep{diaz06}, and deserve 3D simulations in future work.

\acknowledgments
P.F.C. was supported by the Chinese foundation (NSFC 11533005) and Jiangsu 333
Project (No. BRA2017359). Y.Z. supported by China Scholarship Council under
file No. 201606190134. Y.Z. acknowledges Ileyk El Mellah, Jannis Teunissen,
Dimitrios Millas, Tong Shi, Yi-Kang Wang, Kai Yang, and Jie Hong for their help. C.X. thanks FWO
(Research Foundation Flanders) for the award of postdoctoral fellowship. R.K.
was supported by FWO and by KU Leuven Project (No. GOA/2015-014) and by the
Interuniversity Attraction Poles Programme of the Belgian Science Policy Office
(IAP P7/08 CHARM). The simulations were conducted on the VSC (Flemish
Supercomputer Center funded by Hercules foundation and Flemish government) and
on the cluster system in the High Performance Computing Center (HPCC) of
Nanjing University.

\software{MPI-AMRVAC 2.0 \citep{xia18}}

\clearpage

\clearpage
\end{document}